\documentstyle[aps,pra,multicol,epsfig]{revtex}

\newcommand{\be}{\begin{eqnarray}}

\newcommand{\ee}{\end{eqnarray}}
\draft
\tighten

\begin{document}

\title{Laser-induced collective excitations in a two-component 
Fermi gas}

\author{M. Rodriguez$^{1}$ and P. T\"orm\"a$^{2}$}
\address{$^{1}$ Laboratory of Computational Engineering, P.O.Box 9400, 
FIN-02015 Helsinki University of Technology, Finland \\
$^{2}$Department of Physics, University of Jyv\"askyl\"a, P.O.Box 35, 
FIN-40351 Jyv\"askyl\"a, Finland}

\maketitle

\date{\today}

\begin{abstract} 
We consider the linear density response of a two-component (superfluid) Fermi
gas of atoms when the perturbation is caused by laser light. We show that various 
types of laser excitation schemes can be transformed into linear density perturbations,
however, a Bragg spectroscopy scheme is needed for transferring energy and momentum into
a collective mode. This makes other types of laser probing schemes insensitive for 
collective excitations and therefore well suited for the detection of the superfluid
order parameter. We show that for the special case when laser light is coupled between the
two components of the Fermi gas, density response is always absent in a homogeneous system.   
\end{abstract}

\pacs{05.30.Fk, 32.80.-t, 74.25.Gz}

\begin{multicols}{2}[]

\section{Introduction}
Fermionic atoms have already been cooled well below the Fermi temperature
\cite{exp}. Once the quantum regime has been achieved, the predicted BCS 
superfluid transition\cite{prediction} for two different hyperfine
 states with attractive interactions between them would be the next goal to
 achieve. There are several methods proposed for probing the superconducting
 gap \cite{paivi,sorin,light,rest}. Some of them are based on the use of laser 
light. For instance, it has been proposed to measure the 
superfluid coherence and the existence of the superfluid gap by the absorption
of almost on-resonance laser light coupled between the two hyperfine states 
\cite{sorin} or between one of them and another atomic hyperfine state \cite{paivi}. 
Also the use of off-resonant (or very low intensity) laser light \cite{light} 
has been proposed. 
There both hyperfine components are coupled to other excited states
not involved in the pairing, and the excited state populations remain negligible. 

In this paper we review and analyze the linear density response of the 
system to the different laser excitations in the collisionless regime where the 
lifetime of the quasiparticles is much longer than the period of the 
applied field. The motivation for studying the response is two-fold. 
First, in the above-mentioned 
probing schemes for the gap a below-gap density response (e.g. Anderson-
Bogoliubov phonon) would be undesireable because the schemes rely on the 
absence of below-gap excitations. Therefore it is of interest to clarify in 
what kind of laser probing schemes density reponses can be avoided. Second, 
the density response itself may be the object of study and the laser 
excitation schemes that can produce a response are of interest. 

Off-resonant laser excitation caused by two intersecting waves which form
a density grating in space and time (Bragg spectroscopy) has been used to induce 
collective excitations in atomic Bose-Einstein condenstates \cite{meas1}. 
In case of near-resonant laser excitations, we transform
the non-diagonal atom-light interaction Hamiltonian into a form of a density 
perturbation. We show that the initial Hamiltonian of the interacting two-component
Fermi gas is preserved by this transformation 
in case of a contact interaction and therefore
a standard linear density response calculation for a BCS-system can be applied.
In general, an intensity grating in space and time is needed for providing 
momentum and energy transfer for density perturbations. 
We show that when the laser light is coupled
between the two paired hyperfine components, phonons
cannot be excited even when the light forms an intensity grating. 

In section II, the Hamiltonian for a two-component Fermi gas is given and
the density response is defined. The laser probing schemes 
considered in this article are summarized in section III. 
In section IV we transform the perturbation 
caused by the probing lasers into a density perturbation by appropiate
rotation of the coupled states. The results are presented in
detail in section V and summarized in section VI. The linear response
calculation is presented in the Appendix. 

\section{The system}
The Hamiltonian for the two-component Fermi gas, including 
only 2-body interactions between the atoms reads
\be
H_{0} = \int d{\bf r}\sum_{\alpha=\uparrow,\downarrow}
\left[\psi^\dagger_{\alpha}
({\bf r})
\left(-\frac{\nabla^2}{2m}+V_\alpha({\bf r})- \mu_{\alpha}\right)\psi_{\alpha}
({\bf r})\right] \nonumber \\
 + \frac{1}{2}\sum_{\alpha,\beta =\uparrow,\downarrow}
\int d{\bf r}
\int d{\bf r'}{g}_{\alpha \beta}({\bf r}-{\bf r'})
\psi^\dagger_{\alpha}({\bf r})\psi^\dagger_{\beta}({\bf r'})
\psi_{\beta}({\bf r'})\psi_{\alpha}({\bf r}) , \label{ham}
\ee
where the two 
hyperfine states trapped are denoted by $\uparrow$ and $\downarrow$. When the 
interaction ${g}_{\uparrow\downarrow}
({\bf r}-{\bf r'})$ is attractive, the system has been predicted
to undergo a BCS transition \cite{prediction} associated with the appearance
of an order parameter. We assume the system to be below the critical 
temperature and the Hamiltonian $H_0$ treated 
 within the BCS approximation. 
  
The density response of the system to perturbations, caused by e.g. probe
 lasers or
the trapping fields, is studied in the linear response regime \cite{pines}.
The density perturbation is 
a linear function of an external potential $U({\bf r},t)$ expressed through
 the 
density-density response function $\chi({\bf r},t;{\bf r'},t')$:
\begin{equation}
\delta n({\bf r},t)=\delta n_{\uparrow}({\bf r},t)+\delta n_{\downarrow}
({\bf r},t)
=\int d{\bf r}dt\chi({\bf r},t;{\bf r'},t')U({\bf r},t) , 
\end{equation}
where the density operators are defined
 in the usual way $n_{\alpha=\uparrow,\downarrow}({\bf r})=
\psi^\dagger_\alpha({\bf r})\psi_\alpha({\bf r})$. Fourier transforming the
previous expression one obtains 
\begin{equation}
\delta n({\bf k},\omega)=\chi({\bf k},\omega)U({\bf k},\omega).
\end{equation}  
The poles of the reponse function $\chi$ give the collective modes of the system 
while its modulus gives the spectral weight of the modes.

It is known that a perturbation of the form      
$U({\bf r},t)[{\psi}_{\uparrow}^{\dagger}({\bf r}){\psi}_{\uparrow}
({\bf r}) + {\psi}_{\downarrow}^{\dagger}({\bf r})
{\psi}_{\downarrow}({\bf r})]$ leads in the long wavelength limit to 
appearance of the Anderson-Bogoliubov phonon \cite{anna,Grif,AB}, that is, a 
collective mode with energy below the superconducting gap. 

In this paper we study the reponse function $\chi({\bf k},\omega)$ and
potential $U({\bf k},\omega)$ for a perturbation $H'$ created by
different laser probing schemes. Most of the schemes have been
considered in the literature as proposed techniques for observing
the superconducting gap.

\section{Probing schemes}

\subsection{Laser coupling between two states}

{\bf a)} {\it Coupling the two paired states.} Let us assume that the
states $\uparrow$ and
 $\downarrow$ are coupled by light \cite{sorin}.
Using the 
Rotating-Wave-Approximation (RWA) \cite{libros}, the interaction of the laser
 light with 
the matter fields can be described by a time-independent Hamiltonian in which 
the detuning $\delta$ plays the role of a chemical potential and 
a Rabi frequency $\Omega({\bf r})$ characterises the local strength of 
the interaction. The perturbation Hamiltonian is ${H'} =  {H_\mu}+{H}_{T}$,
where the transfer Hamiltonian (${H}_{T}$) and ${H}_{\mu}$ are given 
by 
\be
{H}_{T} = \int d{\bf r}\Omega ({\bf r}){\psi}_{\uparrow}
^{\dagger}({\bf r}){\psi}_{\downarrow}({\bf r}) + \Omega^\ast
({\bf r}){\psi}_{\downarrow}^{\dagger}({\bf r})
{\psi}_{\uparrow}({\bf r}) \nonumber \\
{H_\mu}=\frac{\delta}{2}\int 
d{\bf r}
{\psi}_{\uparrow}^{\dagger}({\bf r}){\psi}_{\uparrow} 
({\bf r})
- \frac{\delta}{2}\int d{\bf r}
{\psi}_{\downarrow}^{\dagger}({\bf r}){\psi}_{\downarrow}
({\bf r}) \label{ppert}
\ee
In principle, one can imagine a direct coupling of the states $\uparrow$ and
$\downarrow$ by one field, in which case $\delta$ and $\Omega$ correspond 
directly to that field. In practice, for coupling of the two hyperfine states
 which have an energy difference far from any laser frequency one may 
prefer a Raman transition. In that case the Rabi frequencies and $\delta$
depend on the parameters from both lasers, especially 
$\Omega \propto \Omega_1^\ast\Omega_2$.    

{\bf b)} {\it Coupling to a non-paired state.} When one of the states
 $\alpha =\{\uparrow$ or 
$\downarrow\}$ is coupled to some excited hyperfine state $e$ \cite{paivi}, the
 perturbation reads
\be
{H}_{T} = \int d{\bf r}\Omega ({\bf r}){\psi}_{\alpha}
^{\dagger}({\bf r}){\psi}_{e}({\bf r}) + \Omega^\ast
({\bf r}){\psi}_{e}^{\dagger}({\bf r})
{\psi}_{\alpha}({\bf r}) \nonumber \\
 H_\mu=\frac{\delta}{2}\int d{\bf r}
{\psi}_{\alpha}^{\dagger}({\bf r}){\psi}_{\alpha} 
({\bf r})
- \frac{\delta}{2}\int d{\bf r}
{\psi}_{e}^{\dagger}({\bf r}){\psi}_{e}
({\bf r}).  \label{excited}
\ee

{\bf c)} {\it Far off-resonant light.}
In \cite{paivi,sorin} the coupled light was proposed to be
almost on-resonant, that is, atomic population is transfered
in the excitation process. In \cite{light} coupling of the
paired states with some excited states was assumed to be
done by far off-resonant or very weak intensity light meaning
the excited state population remains negligible and the
excited state can be eliminated from the problem. 
Far off-resonance coupling can be treated by starting with
an initial Hamiltonian of the same form as (\ref{excited}),
then after adiabatically eliminating the excited
states one arrives at 
$\frac{|\Omega|^2}{\delta} \psi_\alpha^\dagger \psi_\alpha$,
c.f.\ section \ref{nelja}.

\subsection{Space and time dependence of the perturbation}

{\bf d)} {\it Spatial variation of the intensity.} 
The common way of creating intensity gratings used e.g.\  
for creating collective excitations in Bose-Einstein 
condensates \cite{meas1} is to use two intersecting
waves. When two intersecting
 waves with the same polarization but different wave-vector and frequency
are coupled to the same two-level 
system, the total 
Rabi frequency is $ \Omega=\Omega_1e^{i{\bf k_1 \cdot r}-i\omega_1t}+\Omega_2e^
{i{\bf k_2 \cdot r}-i\omega_2 t}$. After making the RWA for the frequency $\omega_1$,
$ \Omega=\Omega_1e^{i{\bf k_1 \cdot r}}+\Omega_2e^
{i{\bf k_2 \cdot r}+i\omega_{12} t}$ where $\omega_{12}=\omega_1-\omega_2$.

Also the beam profile may vary spatially
in the cloud size scale. The Rabi frequency would then be
for instance $\Omega({\bf r}) \propto  e^{-r^2/(2\sigma^2)}$ for a
Gaussian beam shape.  

As will be shown in the following section, the essential feature considering
a possible density response is whether $|\Omega|^2$ is time and space dependent.
Clearly, for Bragg spectroscopy as considered above this is true. 
For the Gaussian beam profile, $|\Omega|^2$ is dependent on position but not
on time. In contrast, for a single laser 
$\Omega = |\Omega|e^{i {\bf k \cdot r} - i \omega t}$
and for a Raman excitation $\Omega \propto |\Omega_1| |\Omega_2 | 
e^{i {\bf k_1 \cdot r} - i \omega_1 t} e^{- i {\bf k_2 \cdot r} + 
i \omega_2 t}$, thus in both cases $|\Omega|^2$
is a constant. 

{\bf e)} {\it Time dependece of the perturbation.} As discussed above, Bragg
scattering schemes lead to a time-varying intensity grating, that is, 
the energy $\omega_{12}$ is transferred to the system via a perturbation that
is proportional to the intensity. 

Another possible source of time-dependence is the turning on of the perturbation. 
As is usually done in the linear response theory \cite{pines}, we consider smooth turning on 
of the perturbing potential.

Direct intensity modulation of the probing laser(s) would be one more
source of time-dependence.

\section{Transforming $H_T$ into a density perturbation}   \label{nelja}

In order to use linear response theory we have to convert all the perturbing 
Hamiltonians of the previous section into a density perturbation. 

{\bf a)} The perturbation term ${H}_{T}$ 
involves products of the field operators of the 
different atomic states. In the mean field approach they correspond to 
the Fock terms $<{\psi}_{\uparrow}^{\dagger}({\bf r})
{\psi}_{\downarrow}({\bf r})>$ which are zero. 
Therefore no linear response theory can be directly applied.

A perturbation of the type (note that the two-component $\psi$ here is
{\it not} the same as that in the standard BCS-theory, used for instance in the
Appendix) 
\be
&& \left(\begin{array}{cc} \psi_\uparrow^\dagger({\bf r}) & 
\psi_\downarrow^\dagger({\bf r})
\end{array} \right)\left(\begin{array}{cc}\delta/2 & \Omega({\bf r})\\
\Omega^\ast({\bf r}) & -\delta/2\end{array} \right)\left(\begin{array}{c}
 \psi_\uparrow({\bf r}) \\ \psi_\downarrow({\bf r})
\end{array} \right) \nonumber \\
&\equiv&  \psi^\dagger ({\bf r}) W \psi({\bf r}) \label{pmatrix}
\ee
can be digonalized for any $\delta$ and $\Omega$ by an appropiate rotation
 in the space of the two states $\uparrow$ and $\downarrow$, 
$\psi=\mathcal{U}$$\tilde{\psi}$
and $\mathcal{U}$=$R_z(\alpha)R_y(\beta)R_z(\gamma)$
with $R_j(\theta)=\exp(i\sigma_j\theta)$, $j=\{x,y,z\}$ and $\sigma_j$ is 
the corresponding Pauli spin matrix. The transformation matrix is
\be
\mathcal{U} =\left(\begin{array}{cc}e^{i(\alpha+\gamma)/2}\cos\beta/2 
&e^{i(\alpha-\gamma)/2}\sin\beta/2\\
-e^{-i(\alpha-\gamma)/2}\sin\beta/2 & 
e^{-i(\alpha+\gamma)/2}\cos\beta/2 \end{array} \right).
\label{trans}
\ee 
The rotated perturbation matrix is $\mathcal{U}^\dagger$W$\mathcal{U}$.
The off-diagonal terms are 
\begin{equation}
e^{-i\gamma}((\delta/2)\sin(\beta) -e^{i\alpha}\Omega^\ast \sin^2(\beta/2)+
\Omega\cos^2(\beta/2)e^{-i\alpha})
\end{equation}
and the complex conjugate. Off-diagonal terms that are zero
 can be obtained by a rotation with the Euler angles $\{\gamma=0, \cos\alpha=
\frac{\Omega+\Omega^\ast}{2|\Omega|},\tan\beta=-\frac{2|\Omega|}{\delta}\}$
 which yields
\be
\left(\begin{array}{cc}
\sqrt{(\frac{\delta}{2})^2+\Omega\Omega^\ast} & 0\\
0 &-\sqrt{\Omega\Omega^\ast+ 
(\frac{\delta}{2})^2}
\end{array}\right)   \label{nelio}
\ee 
for the perturbation matrix. This procedure is extensively used in describing the
interaction of laser light with a two-level system. The form $|\Omega|^2/\delta$
for far off-resonant light comes from (\ref{nelio}) for $|\delta| >> |\Omega|$.  

One has to check whether the rotated states feel the same two-body interaction that
the non-rotated ones, that is, whether the rotated states are still described
by a Hamiltonian of the initial form which implies BCS pairing.
Assuming, for simplicity, that 
$\psi_{\uparrow}$ and $\psi_{\downarrow}$ see the same potential 
$V_\alpha-\mu_\alpha$ , we obtain
${\psi}_{\uparrow}^{\dagger}({\bf r}){\psi}_{\uparrow}
({\bf r}) + {\psi}_{\downarrow}^{\dagger}({\bf r}){\psi}_{\downarrow}
({\bf r}) \rightarrow \tilde{\psi}_{{\uparrow}}^{\dagger}({\bf r})
\tilde{\psi}_{{\uparrow}}({\bf r}) + 
\tilde{\psi}_{{\downarrow}}^{\dagger}({\bf r})\tilde{\psi}_{\downarrow}
({\bf r})$. The kinetic energy term transforms as ${\nabla\psi}_
{\uparrow}^{\dagger}({\bf r}){\nabla\psi}_{\uparrow}
({\bf r}) + {\nabla\psi}_{\downarrow}^{\dagger}({\bf r}){\nabla\psi}_
{\downarrow}
({\bf r}) \rightarrow \nabla\tilde{\psi}_{\uparrow}^{\dagger}({\bf r})
\nabla\tilde{\psi}_{\uparrow}({\bf r}) + 
\nabla\tilde{\psi}_{\downarrow}^{\dagger}({\bf r})
\nabla\tilde{\psi}_{\downarrow}({\bf r})$ if the momentum k
transfered by the laser is small. The extra terms coming in the transformation
of the kinetic energy are $\sim k,k^2$. The momentum of the laser and the 
recoil energy are very small compared to the momentum of the atoms
participating in the process (typically close to the Fermi-surface, $\sim k_F$).

The interaction term in the Hamiltonian transforms
${\psi}_{\uparrow}^{\dagger}({\bf r}){\psi}^\dagger_{\downarrow}
({\bf r'}){\psi}_{\downarrow}({\bf r'}){\psi}_{\uparrow}
({\bf r}) \rightarrow \tilde{\psi}_{\uparrow}^{\dagger} 
\tilde{\psi}_{\downarrow}^\dagger\tilde{\psi}_{\downarrow}
\tilde{\psi}_{\uparrow}$ in case of a contact interaction 
$g({\bf r}-{\bf r'}) \propto \delta ({\bf r} - {\bf r'})$.
As $\tilde{\psi}_{{\alpha}}$ is a 
linear combination of the two species $\uparrow$ and $\downarrow$, one might
expect interaction terms between the same species for the new matter fields
(interactions between both $\tilde{\psi}_{{\uparrow}}$ and
$\tilde{\psi}_{\uparrow}$, and between $\tilde{\psi}_{{\uparrow}}$ and 
$\tilde{\psi}_{\downarrow}$). But the interactions of the type 
$\tilde{\psi}_{\alpha}
\tilde{\psi}_{\alpha}$ are forbidden by the 
fermionic behaviour (commutation rules are preserved by the rotation).
The requirement of a contact interaction is obvious here: $\tilde{\psi}_\alpha ({\bf r})
\tilde{\psi}_\alpha ({\bf r})= 0$, but $\tilde{\psi}_\alpha ({\bf r}) 
\tilde{\psi}_{\alpha} ({\bf r'})$ can be non-zero.
As a summary, since the fermionic field commutator relations are preserved by the
rotation, for contact interaction $\tilde{H}_0= H_0$.
  
We have thus shown that the system can be transformed in such a way that
one can study the collective mode spectra by linear response theory for a
BCS system for a perturbation of the type 
$U (n_\uparrow-n_\downarrow)$, where $U({\bf k},\omega)={\cal F} \left[\sqrt{
\Omega({\bf r}, t) \Omega({\bf r}, t)^\ast+(\frac{\delta}{2})^2}\right]$, and the 
rest of the Hamiltonian still corresponds to the BCS Hamiltonian. The main 
difference to the standard linear density response treatment for an BCS system
 is the minus sign between $n_\uparrow$ and $n_\downarrow$. We will show that
this leads to zero overall response in the homogeneous case.

{\bf b)} When diagonalizing the interaction Hamiltonian of Eq.(\ref{excited})
in the same way as in {\bf a)}, the initial Hamiltonian $H_0$ is {\it not} 
preserved. 
We thus obtain a perturbation $U(n_\alpha - n_e)$ with $U({\bf k},\omega)$ as
given above, but with the initial Hamiltonian modified, $\tilde{H}_0 \neq
H_0$.

{\bf c)} In the case of coupling to an excited state by far-detuned light 
the rotation does not modify the initial Hamiltonian because at the 
limit $|\delta| >> |\Omega|$ the transformed states are very close to 
the initial ones, $\psi_\alpha \sim \tilde{\psi}_\alpha$ and 
$\psi_e \sim \tilde{\psi}_e$, moreover the excited state population
is assumed to be negligible. Thus the light-atom interaction Hamiltonian is 
in the form of a density perturbation, and the initial Hamiltonian is
preserved. 

In all these cases, $U({\bf k},\omega)$ can be either a constant $U({\bf k},
\omega)= U'$ or dependent on ${\bf k}$ and $\omega$. The latter happens 
when the intensity $|\Omega|^2$ is space and time dependent. Note that the
$\textbf{k}$-dependence $\Omega e^{i {\bf k \cdot r}}$ is not sufficient, 
that is, 
the laser momentum given in a single-laser or Raman two-level excitation 
(near-resonant or off-resonant) is not enough to 
transfer momentum through a density perturbation.

\section{Results}

\subsection{GRPA for the case of light coupled between the two paired states}
The response of the system to the perturbation $U(k,\omega) 
(n_\uparrow-n_\downarrow)$ in a homogenous system is calculated in the 
Appendix following the derivation in\cite{Grif}. 
The result is that for the homogeneous case, the responses of two components 
are equal but with opposite signs and no 
collective modes are excited at any frequencies. 

The response of one of the spin 
components is $L_{11}=\frac{(a+b)(1+g_0R)}{(1-g_0R)(g_0+a+b)}$, where
we use the 
integrals a, b, c, d, R as defined in \cite{Grif}. The reponse 
of the spin component has   
the factor (a+b) which at T=0 in the low q,$\omega$ region is 
$a+b \approx -N(\epsilon_F)c_B^2q^2/(12\Delta^2)$ where $N(\epsilon_F)$ is the
 density of states at the Fermi surface and the sound velocity 
$c_B^2=v_F^2/3$. This factor
 is very small and it makes the response for one spin component
 negligible even without the cancellation effect.
This is because the superfluid cannot participate in the relative motion of 
the two spin components\cite{Legget}.
Above $T_c$, b vanishes and the 
response of one of the spins is not negligible anymore, 
as the integral $a$ is some fraction of $N(\epsilon_F)$. However, due to the 
cancellation the overall response is zero. 

We have considered the homogeneous case, that is, when the trapping potential
treated within the local density approximation is sufficient.
The strongly trapped case can be approached by using multipole expansions \cite{BruunM}:
the overall response
is not zero for spin-dipole excitations in the trapped case
above $T_c$. The difference to our results is a 
consequence of the geometry, as in the homogeneous case we get the same 
spatially uniform response in opposite directions for both spin species. 
As $T \rightarrow 0$, one gets vanishing response also for the 
inhomogeneous case \cite{BruunM}, at least at energies below the gap, 
because one needs to break pairs in order to have relative motion of the two
 components.

\subsection{Density response for different probing schemes}

{\bf a)}{\it Coupling the paired states.}
The perturbation reads $U(n_\uparrow-n_\downarrow)$ for the rotated states, 
where $U({\bf k},\omega)= 
{\cal F}[\sqrt{|\Omega(\textbf{r},t)|^2+\delta^2}]$. When the light is far 
off-resonant, one can take the limit $U({\bf k},\omega) \sim {\cal F}[|\Omega({\bf r}, t)|^2/
\delta]$. In the case of homogeneous laser intensity $\Omega=
|\Omega| e^{i{\bf k \cdot r}}$, $U \propto \delta(\omega)\delta(\textbf{k})$. 
There is no 
density reponse simply because U does not provide the momentum and energy
 for a collective excitation. 

A different behaviour arises when using a Bragg scattering scheme to
provide an intensity grating. Assuming $|\Omega_1|=|\Omega_2|$ ($\Omega_1, \Omega_2$ as introduced in section III d)) one obtains 
$|\Omega|^2 = |\Omega_1|^2(2+2 \cos(\omega_{12} t- {\bf k_{12} \cdot r}))
 \sim |\Omega_1|^2 2\cos(\omega_{12}t- {\bf k_{12} \cdot r})$ 
and the density response $\chi(\omega, {\bf k})$
has to be analyzed at 
$\chi(\omega_{12}, {\bf k_{12}})$. The density response $\chi$ gives zero 
in the homogeneous case for all temperatures, as one 
component annihilates the response of the other, and for the inhomogeneous 
case well below $T_c$ \cite{BruunM} as considered in the previous subsection.

For a general case 
where the beam profile of the laser light gives a spatially varying 
intensity of the form $|\Omega({\bf r})|^2 \sim e^{\frac{r^2}{\sigma^2}}$,
 the far off-resonant perturbation reads
$U({\bf k},\omega)=\frac{|\Omega(k)|^2}{\delta} \sim 
e^{\frac{\sigma^2k^2}{4}}$. If $\sigma$ is some fraction $f$ of the trap size 
$R_{TF}$, one can estimate the momentum given by $p=\frac{\hbar}{\sigma}$. A 
collective mode with an energy of the order of the gap energy (e.g.\ half 
the gap) has momentum 
$q \sim \sqrt{3}\frac{\Delta_G}{v_F}$ \cite{pines,anna} and the ratio between 
them is $\frac{p}{q}=\frac{\hbar\omega_T}{f\Delta_G}$. For typical 
parameters, this
could be a non-negligible number, however, this alone is not enough to 
produce a density response because of the lack of the time dependence, 
i.e.\ $U(\omega, {\bf k}) \propto \delta(\omega) f({\bf k})$.   

{\bf b)} {\it Coupling to an excited state.} When diagonalizing the
interaction Hamiltonian in this case the initial Hamiltonian was
modified. Therefore the linear response calculation for a BCS system
cannot be directly applied. One may guess, however, that the 
arguments about the perturbation $U$ being time- and space-dependent
will hold also in this case even when the response function $\chi$
would have a different form. Therefore we do not expect a density
response except when a Bragg spectroscopy scheme is used. 

{\bf c)} {\it Far off-resonant coupling.} 
When both paired states are coupled to some excited states and the light is 
far detuned, the density perturbation potential is proportional to 
$|\Omega|^2/\delta$. No collective mode of the system is excited if there is no 
intensity variation in space and time. When considering intensity modulations of the type 
$|\Omega|^2\sim |\Omega_1|^2 2\cos(\omega t-{\bf k \cdot r})$, an
 Anderson-Bogoliubov phonon can be excited \cite{anna,Grif,AB}.

\section{Conclusions}

We have reviewed the density response of some typical laser excitation
schemes. We have shown that since the BCS Hamiltonian for a contact interaction
is preserved under a rotation, most of the considered laser excitations can be
expressed in terms of a perturbation acting on the density. In this form,
the perturbation potential is proportional to $|\Omega|^2$ where $\Omega$
is the (effective) Rabi frequency. Therefore, even when the laser light provides
momentum and energy ($\Omega \propto e^{i {\bf k_L \cdot r}} e^{i \omega_L t}$),
the transformed potential acting as a density perturbation is not time- and
space-dependent. This leads to absence of a density response whenever
$|\Omega|^2$ is a constant spatially and temporally. 
This makes many proposed laser-probing schemes
well suited for observing the superconducting gap since they do not induce below-gap
collective excitations.

For Bragg scattering, $|\Omega|^2 \propto f({\bf r},t)$. In this case
Anderson-Bogoliubov phonons can be excited, in general. The exception is the case
when the laser(s) couple between the two paired components of the gas. We have
shown that, in the homogeneous case, the density response becomes zero
because the contributions of the components cancel each other. In a harmonic
trap, spin-dipole response is predicted for temperatures near $T_c$. 
Therefore, the presence or absence of low-energy collective excitations
under a perturbation of the type $U(\omega,k)(n_\uparrow - n_\downarrow)$
could be used to observe whether the trapped system can be approximated
by a homogeneous system (local density approximation) or whether the
trapping effects are dominant.

\section{Appendix}
The formalism used is based on the functional differentiation
 technique as described in \cite{KadBaym} following the derivation in 
\cite{Grif}. The response function $\chi$
is viewed as a functional derivative of the one-particle matrix Green's 
functions $G$ with respect to the external field $U({\bf r},t)$. 
To allow for pairing in the superconducting phase, it is convenient to work 
with a single-particle Green's function G given by a $2\times2$ matrix 
defined as
\be
G(1,2) \equiv  - \langle T\Psi(1)\Psi^\dagger(2)\rangle=
\left(\begin{array}{cc}G_\uparrow(1,2) & F(1,2) \\F^\ast(1,2) & 
-G_\downarrow(2,1) \end{array}\right),  
\ee 
where $\Psi = (\psi_\uparrow \quad \psi_\downarrow^\dagger)^T$ and
$1\equiv({\bf r_1},\tau_1)$.
Imaginary times (Matsubara formalism) are used so that one can deal 
with finite temperatures. In the absence of external fields, the equal time
($\tau_2=\tau_1^+$) single-particle Green's function components reduce to
$F({\bf r})\equiv \langle \psi_\downarrow({\bf r})\psi_\uparrow({\bf r})
\rangle$, the s-wave order parameter, and 
$G_{\uparrow,\downarrow}({\bf r})=\langle n_{\uparrow,\downarrow}
({\bf r})\rangle$.

>From the equation of motion of the Green's
function one gets the generalized Dyson equation for G(1,2) in terms of
$G_0$, the non-interacting single-particle Green's function, the matrix 
self-energy  $\Sigma(1,2)$ which is evaluated in the pairing approximation 
(Hartree-Fock-Bogoliubov), and W(1) which is the  
the external perturbing field matrix. 

The density response 
matrix is obtained in RPA by taking the functional derivative of 
the Green's function 
with respect to the external field U. One can define the three-point
 correlation
function $L(1,2,5)\equiv -\sigma_3 \frac{\delta G(1,2)}{\delta U(5)}$, whose
limit $L(1,2)\equiv L(1,1^+,2)$ will give the density-density response 
function $\chi(1,2)=L_{11}(1,2)+L_{22}(1,2)$. Here $\sigma_3$ is the third
 Pauli matrix.

When deriving the Dyson equation for G(1,2) with respect to U in order to get
 the three-point correlation function L, both the self energy matrix and the 
external field matrix W contribute. The lowest-order 
(single-bubble) result \cite{Grif} for $L$ is given by
\begin{equation} 
L^0(1,2,5)=-\sigma_3\int d\bar{3}\int
 d\bar{4}G(1,\bar{3})\frac{\delta W(\bar{3},\bar{4})}{\delta U(5)}
G(\bar{4},2).  \label{sbubble}
\end{equation}
For the Anderson-Bogoliubov 
phonon (see \cite{anna,Grif}) the pertubation matrix $W(1)=U(1)\sigma_3$, 
and $L_{AB}^0(1,2,5)=\tilde{G}(1,5)\tilde{G}(5,2)$,  where we have 
introduced $\tilde{G}\equiv\sigma_3G$. In our case 
$L^0(1,2,5)=\tilde{G}(1,5){G}(5,2)$ because the perturbation matrix
$W(1)=U(1){\mathcal I}$ due to the minus sign in $U(n_\uparrow - n_\downarrow)$.
>From now on, we denote by the subindex AB the quantities that are 
calculated for the perturbation matrix $W(1)=U(1)\sigma_3$.  

It is useful to rewrite the 
GRPA integral equations in terms of irreducible two-particle 
Green's functions
$\bar{L}_{ij}(1,2,5)$, and for the  
homogeneous system, to Fourier transform in 
order to solve coupled equations. The result reads \cite{Grif}
\begin{equation}
L_{ij}({\bf q},i\omega_n)=\bar{L}_{ij}({\bf q},i\omega_n)+
\frac{\bar{L}_{ABij}({\bf q},i\omega_n)
g_{\uparrow \downarrow}({\bf q})\bar{L}_{ll}({\bf q},i\omega_n)}
{1-g_{\uparrow \downarrow}({\bf q})\bar{L}_{ABll}({\bf q},i\omega_n)},
\end{equation}
where equal indices mean summation over the possible values. We denote by
 the subindex AB the three-point correlation functions for the 
Anderson-Bogoliubov phonon type of perturbation.

The sum of diagonal terms reduces to
\be
\chi({\bf q}, i\omega_n)&=& L_{11}({\bf q},i\omega_n)+L_{22}({\bf q},i\omega_n) 
\nonumber \\ &=&
\frac{\bar{L}_{ii}({\bf q},i\omega_n)}
{1-g_{\uparrow \downarrow}({\bf q})\bar{L}_{ABll}({\bf q}, i\omega_n)} .
 \label{imp}
\ee
This shows that all linear perturbations have the same poles 
as the Anderson-Bogoliubov phonon (plus possibly some additional ones), 
 but the spectral weight depends on the trace of the irreducible
 two-particle Green's
functions $\bar{L}_{ii}({\bf q},i\omega_n)$ of the specific perturbation. 

For a contact interaction ($g_{\uparrow \downarrow}({\bf q})=g_0$), 
the equation for the irreducible correlation function \cite{Grif} reduces to a 
set of linear algebraic equations   
\begin{equation}
\bar{L}_{ij}({\bf q},i\omega_n)={L}^0_{ij}({\bf q},i\omega_n)-
L^0_{ABiklj}({\bf q},i\omega_n)g_0\bar{L}_{kl}({\bf q},i\omega_n), \label{a}
\end{equation}
where we have defined the four-index tensor $L^0_{ABiklj}$, indicating 
components of the two factor matrices.
Defining column vectors $\bar{L}=\left(\begin{array}{c}\bar{L}_{11}\ 
\bar{L}_{12}\ \bar{L}_{21}\ \bar{L}_{22}\end{array}\right)^T$ and a 
$4\times4$ matrix $L^0_{(AB)mn}$ as in
\cite{Grif}, Eq.(\ref{a}) reduces to
\begin{equation}
\bar{L}=[I+g_0L^0_{AB}]^{-1}L^0. \label{util}
\end{equation}
As discussed in \cite{Grif} one can get $L^0_{ijkl}$ by Matsubara frequency
 summations, (Eq.(4.27) in \cite{Grif}) 
and using the symmetry properties then reduce just to 6 independent elements.
 For the weak-coupling limit and contact interaction, the independent elements
 are reduced to four: a, b, c, d which are integrals defined in
 \cite{Grif} (Eq.A8). $L^0_{m}$ can 
be calculated from the $4\times 4$ matrix by 
$L^{0T}_{m}=\left(\begin{array}{c}
L^0_{1ll1}\ L^0_{1ll2}\ L^0_{2ll1}\ L^0_{2ll2} \end{array}\right)$.

Now we derive the $4\times 4$ matrix for our case when 
$W(1)=U \mathcal{I}$ by reconsidering the symmetry properties of the matrix
 elements 
in order to derive the vector $L^0_{m}$ that we insert in Eq.(\ref{util}). 
The sum of the first and last component of $\bar{L}_m$ contribute to 
Eq.(\ref{imp}) and characterize the spectral weight and
 perhaps some additional poles of the response function. Such a perturbation 
gives from (\ref{sbubble}) the 
lowest order correlation function $L^0(1,2,5)=\tilde{G}(1,5)G(5,2)$ (c.f.
$L_{AB}^0(1,2,5)=\tilde{G}(1,5)\tilde{G}(5,2)$). We 
 calculate the
 $4\times4$ matrix $L^0_{mn}$ using the symmetry properties of the 
new $L^0_{ijkl}$ elements, in the same fashion as in $\cite{Grif}$ (see 
appendix A there) 
obtaining
\be
L^0_{mn}&=&\left(\begin{array}{cccc}a & c & -c &b\\ c& -d&b&-c\\
c&-b&d&-c\\-b&c&-c&-a\end{array}\right).
\ee  
It leads to $L_m^{0T}=\left(\begin{array}{cccc}a+b & 0 & 0& -a-b
\end{array}\right)$. Multiplying $L_m^0$ by $[I+g_0L^0_{ABmn}]^{-1}$ as in
(\ref{util}) one gets $\bar{L}_m^{T}=\left(\begin{array}{cccc}
\frac{a+b}{g_0+a+b} & 0 & 0& 
-\frac{a+b}{g_0+a+b}
\end{array}\right)$. This means $\bar{L}_{11}=-\bar{L}_{22}$, and the 
spectral weight in (\ref{imp}) vanishes and $\chi=0$.

Inserting the well-known result \cite{Grif,anna}
 $\bar{L}_{ABll}=\frac{2R}{1+g_0R}$ into Eq.(\ref{imp}), one obtains the spin 
density response $L_{11}=\frac{(a+b)(1+g_0R)}{(1-g_0R)(g_0+a+b)}$, where
 $R=A+\frac{4c^2}{1+g_0B}$ and $B\equiv b+d$, $A\equiv a-b$ are defined 
in \cite{Grif}.

In the case of a general perturbation
\begin{equation}
W=U\left(\begin{array}{cc}1 & 0\\0 & c_u \end{array}\right),
\end{equation} 
where $c_u$ is any real number, the single-bubble result is 
\begin{equation}
L^0_{ijkl}=\tilde{G}_{ij}\left(\left(\begin{array}{cc}1 & 0\\0 & c_u 
\end{array}\right)_{ks}G\right)_{sl}.
\end{equation}
By calculating the symmetry properties of the elements of $L_{mn}$, one 
obtains $L_m^{0T}=\left(\begin{array}{cccc}a+c_ub & c-c_uc & c-c_uc & 
-c_ua-b\end{array}\right)$ which leads to
\begin{equation}
\bar{L}_{11}-\bar{L}_{22}=\frac{(c_u+1)(a+b)}{1+g_0(a+b)}
\end{equation}
and
\begin{equation}
\bar{L}_{11}+\bar{L}_{22}=\frac{(1-c_u)R}{1+g_0R}.
\end{equation}
The reponse for one spin would be
\begin{equation}
L_{11}=\frac{r_u}{(1+g_0R)(1+g_0B)}
\end{equation}
where $r_u=g_0AB+g_02c^2\frac{(1-c_u)+2g_0(a+b)}{1+g_0(a+b)}+
\frac{(a+c_ub)+g_0(a^2+c_ub^2+db(1+c_u))}{1+g_0(a+b)}$.

\end{multicols}

\begin{thebibliography}{99}
\bibitem{exp}
B.\ DeMarco, S.\ B.\ Papp, and D.\ S.\ Jin, Phys. Rev. Lett. \textbf{86},
5409 (2001); B.\ DeMarco and
D.\ S.\ Jin, Science {\bf 285} 1703 (1999); A.\ G.\ Truscott, K.\ E.\ 
Strecker, W.\ I.\ McAlexander, G.\ B.\ Partridge and R.\ G.\ Hulet, 
Science {\bf 291}, 
2570 (2001) ; F.\ Schreck. L.\ Khaykovich, K.\ L.\ Corwin, G.\ Ferrari, T.\
 Bourdel, J.\ Cubizolles and C.\ Salomon, Phys.\ Rev.\ Lett.\ \textbf{87}, 
080403 (2001); Z.\ Hadzibabic, C.\ A.\ Stan, K.\ Dieckmann, S.\ Gupta, M.\ 
W.\ Zwierlein, A.\ G\"orlitz and W.\ Ketterle, cond-mat/0112425;
S.\ R.\ Granade, M.\ E.\ Gehm, K.\ M.\ O'Hara and J.\ E.\ Thomas, 
cond-mat/0111344.
\bibitem{prediction}
H. T. C. Stoof, M. Houbiers, C. A. Sackett, and R. G. Hulet,
Phys. Rev. Lett. {\bf 76}, 10 (1996); M. Holland, S. J. J. M. F. 
Kokkelmans, M. L. Chiofalo, and R. Walser, Phys. Rev. Lett. \textbf{87}, 
120406 (2001); L.\ Viverit, S.\ Giorgini, L.P.\ Pitaevskii,
and S.\ Stringari, Phys. Rev. A \textbf{63}, 033603 (2001); H. Heiselberg,
C.\ J.\ Pethick, H.\ Smith and L.\ Viverit, Phys. Rev. Lett. \textbf{85}, 
2418 (2000).
\bibitem{paivi} 
 P. T\"orm\"a and P. Zoller, Phys. Rev. Lett. {\bf 85}, 487 
(2000); G. M. Bruun, P. T\"orm\"a, M. Rodriguez, and P. Zoller, Phys. 
Rev. A {\bf 64}, 033609 (2001).
\bibitem{sorin}
Gh.-S. Paraoanu, M. Rodriguez and P. T\"orm\"a, J.\ Phys.\ B \textbf{34}, 
4763 (2001).
\bibitem{light}
W.\ Zhang C. A. Sackett, and R. G. Hulet, Phys.\ Rev.\ A \textbf{60},
 504 (1999); 
J.\ Ruostekoski, Phys.\ Rev.\ A \textbf{60}, R1775 (1999), J.\ Ruostekoski,
Phys.\ Rev.\ A {\bf 61}, 033605 (2000); 
F.\ Weig and W.\ Zwerger, Europhys.\ Lett.\ \textbf{49}, 282 (2000).
\bibitem{rest}
M.\ A.\ Baranov and D.\ S.\ Petrov, Phys.\ Rev.\ A \textbf{62}, 041601(R) 
(2000); M.\ Farine {\it et al}, Phys.\ Rev.\ A \textbf{62}, 013608 (2000); 
G.\ M.\ Bruun and C.\ W.\ Clark, J.\ Phys.\ B \textbf{33}, 3953 (2000).
\bibitem{meas1}
D.\ M.\ Stamper-Kurn, A.\ P.\ Chikkatur, A.\ G\"orlitz, S.\ Innoye, S.\ 
Gupta, D.\ E.\ Pritchard, and W.\ Ketterle, Phys.\ Rev.\ Lett.\ {\bf 83},
2876 (1999).
\bibitem{pines}
D.\ Pines and P.\ Nozi\`eres, {\it The Theory of Quantum Liquids: Vol I}
(Benjamin, New York, 1966)  
\bibitem{anna}
A.\ Minguzzi, G.\ Ferrari and Y.\ Castin, Eur.\ Phys.\ J.\ D, \textbf{17}, 49
(2001).
\bibitem{Grif}
R. C\^ot\'e and  A. Griffin, Phys.\ Rev.\ B \textbf{48}, 10404 (1993).
\bibitem{AB} 
P.\ W.\ Anderson, Phys.\ Rev.\textbf{112}, 1900 (1958); N.\ N.\ Bogoliubov, V.\
V.\ Tolmachev and D.\ V.\ Shirkov, {\it A New Method in the Theory of 
Superconductivity} (Academy of Science, Moscow 1958, New York 1959).
\bibitem{libros} 
B.\ W.\ Shore, {\it The theory of coherent atomic excitation}
(Wiley, New York, 1990). Vols. I and II.
\bibitem{KadBaym}
L. P. Kadanoff and G. Baym, {\it Quantum Statistical Mechanics} (Benjamin,
New York, 1962).
\bibitem{Legget}
A.\ J.\ Legget, Phys. Rev. \textbf{147}, 119 (1966).
\bibitem{BruunM}
G.\ M.\ Bruun and B.\ R.\ Mottelson, Phys.\ Rev.\ Lett. \textbf{87},
 270403 (2001).
\end{thebibliography}
\end{document}